\documentclass[11pt,twoside]{article}

\usepackage{amsmath}
\usepackage{amsfonts}
\usepackage{amsthm}
\usepackage{graphicx}
\usepackage{caption}
\usepackage{subcaption}
\usepackage{setspace}
\usepackage{epstopdf}
\usepackage{wrapfig}
\usepackage{fancyhdr}
\usepackage{geometry}

\theoremstyle{definition}

\fancypagestyle{mystyle}{
\fancyhf{}
\fancyhead[OC]{\small\it J. M. Taylor}
\fancyhead[EC]{\small\it Non-Gaussian chain statistics in LC elastomers.}
\fancyfoot[C]{\thepage}
}

\fancypagestyle{first}{
\fancyhf{}
\fancyfoot[C]{\hrulefill\\
{\small 1. Address Correspondence to Jamie M. Taylor, Mathematical Institute, Woodstock Road, Oxford, OX2 6GG, UK. Email: jamie.taylor@maths.ox.ac.uk}\\\vspace{0.1in} \thepage}
}

\author{{\bf Jamie M. Taylor}\\ Mathematical Institute, University of Oxford}
\title{\bf Non-Gaussian chain statistics and finite extensibility in liquid crystal elastomers}
\date{}

\begin{document}
\pagestyle{mystyle}
\doublespace
\maketitle
\thispagestyle{first}
\abstract{In this work we will derive an anisotropic generalisation of the finitely extensible chain model, due to Kuhn and Gr\"un, which is well known in rubber elasticity. This provides a chain energy that couples elastic behaviour to a probability distribution describing the orientations of liquid crystal monomers within a main chain elastomer. The key point is to invoke a maximum relative entropy assumption on the distribution of bond angles in an observed chain. The chain energy's fourth order Taylor expansion is also given, which couples to the second and fourth moments of the nematic distribution function only.}

\section{Introduction}

The Neo-Hookean free energy due to Warner, Bladon and Terentjev \cite{bladon1993transitions} is a highly favoured free-energy density for describing elastic deformations in liquid crystal elastomers. Whilst it has been effective in predicting many observed properties of liquid crystal elastomers, for example soft modes \cite{warner1994soft} and stripe domain instabilities \cite{verwey1996elastic}, its derivation, based on Gaussian chain statistics, relies on the assumption that all chains in the polymer network are far away from their maximal extension, which limits the validity of the model in the large-strain regime. Attempts to overcome these issues have included an anisotropic analogue of the phenomenological Ogden model \cite{agostiniani2012ogden}, and an argument based on finite extensibility effects that produces a correction to the Neo-Hookean energy \cite{mao1998finite}.

In the case of isotropic elasticity, there are a wide variety of models (see e.g. \cite{boyce2000constitutive}) to describe the large-strain behaviour of rubbers that are fundamentally based on a description due to Kuhn and Gr\"un of the energy of a single polymer chain within a cross-linked network \cite{kuhn1942beziehungen}. Loosely speaking, their chain energy corresponds to a maximum entropy assumption on an observed end-to-end vector, spanning the crosslinking points of a chain, and has the key consequence that it requires an infinite amount of energy to reach the taut-chain limit. The purpose of this work is to generalise the model of Kuhn and Gr\"un, for brevity denoted the KG-model, to permit anisotropic behaviour. In particular the free energy will be coupled to the orientation distribution function that describes the alignment of liquid crystal molecules within a main-chain liquid crystal elastomer. The derivation presented in this work mainly follows that of the KG-model, with the key difference that the {\it a priori} probability of finding a monomer with a particular orientation in a free system will no longer be assumed to be isotropic, and instead will be coupled with the nematic orientation distribution function. The derivation is then shown to be equivalent to a maximum relative entropy assumption on the end-to-end vectors of the chains, where the relative entropy is taken against a probability distribution that reflects the anisotropic nature of the elastomer. 

In order to describe the energy of deformation for the material one must make assumptions on both the behaviour of the polymer network in the reference configuration, as well as the manner in which it deforms with the material. By considering a second order Taylor approximation of the chain energy and taking an affine displacement assumption on the chains, the deformation energy will reduce to the the Neo-Hookean formulation. Furthermore the fourth order Taylor expansion can be taken to produce a deformation energy that couples to the second and fourth order moments of the nematic distribution in the both reference and deformed configurations. This can be viewed as either a correction to the Neo-Hookean theory, or an approximation of a full-network theory. The paper is then concluded by showing some numerically obtained results to illustrate the behaviour of the energy. 

\section{Foundations}

\subsection*{Assumptions}

Consider a chain in the crosslinked polymer network, with $R \in \mathbb{R}^3$ spanning the two points of cross-linking, the so-called end-to-end vector. As in \cite{kuhn1942beziehungen}, the methodology will be first to consider a discrete system, where the monomers can have one of finitely many orientations. Then the entropically optimal configuration for a given $R$ will be found in this discrete case, and continuum limits of the solution will be taken. 

Partition the surface of the unit sphere into $M$ disjoint sections, $(U_j)_{j=1}^M$ so that $\bigcup\limits_{j=1}^M U_j = \mathbb{S}^2$, and $U_i \cap U_j = \emptyset$ for $i \neq j$. Furthermore assume that the maximum diameter of $(U_j)_{j=1}^M$ tends to zero as $M\to\infty$. We assume that the admissible orientations $p_j$ for $j=1,...,M$ satisfy $p_j \in U_j$. Let $n_j$ denote the average number of monomers with orientation $p_j$. We make the following assumptions on the chain.
\begin{enumerate}
\item \label{HNum}The total number of monomers is equal to $N$, that $\sum\limits_{j=1}^M n_j=N$.
\item \label{HR}The length of each monomer in the chain is equal to $\ell>0$, so that $\ell\sum\limits_{j=1}^M n_jp_j=R$.
\end{enumerate}
These two assumptions are common between this work and the work of Kuhn and Gr\"un. The next assumption is to establish the description of the rod-like liquid crystal and flexible components of the polymer chain. 
\begin{enumerate}
\setcounter{enumi}{2}
\item \label{HSplit} The monomers can be split into two classes, those in the flexible polymer backbone of number fraction $1-\gamma$, and the nematic mesogens of number fraction $\gamma$.
\end{enumerate}
The next assumption is the key deviation from the Kuhn and Gr\"un model, so that the entropy is taken with respect to an anisotropic distribution, which will drive the key properties of our model. It makes an assumption on what Treloar referred to as the ``{\it a priori}" probability of finding a monomer with a certain orientation \cite{treloar1975physics}, and can loosely be interpreted as the natural state of the monomers in the absence of the rest of the polymer network. Here the notation $\mathcal{P}(\mathbb{S}^2)$ denotes the set of probability distributions on the sphere which are absolutely continuous with respect to the Hausdorff measure. 
\begin{enumerate}
\setcounter{enumi}{3}
\item \label{HPriori} The {\it a priori} probability of finding a mesogenic monomer with a particular orientation is described by some given $\rho \in \mathcal{P}(\mathbb{S}^2)$, and the {\it a priori} probability of finding a backbone monomer with a given orientation is described by the uniform distribution $\frac{1}{4\pi}$.
\end{enumerate}

The law of total probability then gives that the {\it a priori} probability of finding an arbitrary monomer with a particular orientation is given by 
\begin{equation}\label{eqRhoT}
\rho^T=\frac{1-\gamma}{4\pi}+\gamma \rho.
\end{equation}

There are generally a large number of chain configurations that could provide the observed end-to-end vector, so we need an assumption in order to determine what the most likely chain configuration is.
\begin{enumerate}
\setcounter{enumi}{4}
\item \label{HEntropy}(Maximum entropy assumption) If the end-to-end vector $R$ is observed, then the corresponding $n_j$ are of maximum entropy, given the {\it a priori} distribution $\rho^T$.
\end{enumerate}
In the absence of an internal energy term, which will be taken in most of this work, the free energy depends only on the entropy and temperature. In particular, maximising the entropy is equivalent to minimising the free energy, and the free energy can be thought of as being purely entropic. 

\subsection*{Derivation}

Let $\Delta p_j$ denote the surface area of $U_j$. Define the probability of a monomer having orientation $p_j$ to be $\rho^T(p_j)\Delta p_j$, corresponding to the approximation
$$\int_{U_j} \rho^T(p) \, dp \approx \int_{U_j} \rho^T(p_j) \, dp = \rho^T(p_j)\Delta p_j.$$
Treating the orientations of each monomer as independent and using the multinomial distribution, the probability of the observed chain $R$ is given by 
$$W=N! \left(\prod\limits_{j=1}^M (n_j!)\right)^{-1}\prod\limits_{j=1}^M (\rho^T(p_j)\Delta p_j)^{n_j}$$
It should be remarked that in the original derivation of Kuhn and Gr\"un, they implicitly assumed an isotropic {\it a priori} distribution in the previous step whereas we take Assumption (\ref{HPriori}) here. Much of the remainder of the derivation is common to their work. Taking negative logarithms (to give an entropic energy) and using Stirling's approximation ($\ln n! \approx n\ln n - n$), approximately it holds that
$$-\ln W \approx N-N \ln N + \sum\limits_{j=1}^M \left(n_j \ln n_j - n_j\right) - \sum\limits_{j=1}^M n_j \ln \left(\rho^T(p_j)\Delta p_j\right)$$
Using Assumption (\ref{HEntropy}) so that $(n_j)$ must be energetically optimal, and using Lagrange multipliers $\alpha \in \mathbb{R}$, $\beta \in\mathbb{R}^k$ to satisfy the constraints from Assumptions (\ref{HNum},\ref{HR}),
\begin{displaymath}
\begin{split}
0=&\frac{\partial }{\partial n_j} \left(-\ln W- \alpha \left(\sum_{i=1}^M n_i-N\right)- \beta \cdot \left(\sum\limits_{i=1}^M  n_ip_i-\frac{1}{\ell} R\right)\right)\\ =&\ln n_j -\ln \left(\rho(p^T_j)\Delta p_j\right) - \alpha - \beta \cdot p_j. 
\end{split}
\end{displaymath}
Rearranging then gives
$$n_j=\exp(\alpha+\beta \cdot p_j)\rho^T(p_j)\Delta p_j.$$
Substituting this back into the energy gives 
\begin{displaymath}
\begin{split}
-\ln W =& N-N\ln N+ \sum\limits_{j=1}^M n_j \left(\alpha+ \beta \cdot p_j-\ln \left(\rho^T(p_j)\Delta p_j\right)\right)-\sum\limits_{j=1}^M n_j \ln \left(\rho^T(p_j)\Delta p_j\right)\\
=& N-N\ln N+N\alpha+\beta \cdot \frac{1}{\ell} R\\
=& N-N\ln N +N\left(\alpha+\beta\cdot\frac{1}{N\ell}R\right).
\end{split}
\end{displaymath}

Taking the continuum limit as $M \to \infty$, a measure $\xi$ is obtained (depending on $R$ and $\rho$) of bond angles so that 
\begin{equation}\label{formOfSigma}
d\xi(p) = \exp(\alpha + \beta \cdot p) \rho^T(p) dp
\end{equation}
with the requirements that 
\begin{displaymath}\begin{split}
\int_{\mathbb{S}^2}d\xi(p) =& \int_{\mathbb{S}^2} \exp(\alpha + \beta \cdot p )\rho^T(p) \, dp  = N,\\
\\\int_{\mathbb{S}^2} p \, d\xi(p) =& \int_{\mathbb{S}^2} \exp(\alpha + \beta \cdot p ) \rho^T(p) p \, dp = \frac{1}{\ell}R
\end{split}
\end{displaymath}
The constant $\alpha$ can be eliminated to obtain an implicit representation of $\beta$ by observing that
$$\frac{1}{N \ell}R = \frac{ 1}{\int_{\mathbb{S}^2} \exp(\beta\cdot p ) \rho^T(p) \, dp}\int_{\mathbb{S}^2}p \exp(\beta \cdot p ) \rho^T(p) \, dp.$$
In particular this gives $\frac{1}{N\ell}R$ as the first moment of a continuous probability distribution on the sphere. Therefore we must have that $|R|<N\ell$, or otherwise there cannot be solutions $\beta$.

Let $\sigma(p)=\frac{1}{Z}\exp(\beta \cdot p)\rho^T(p)$ be the minimising distribution probability distribution, so $\int_{\mathbb{S}^2}\sigma(p)\,dp=1$ and $\int_{\mathbb{S}^2}p\sigma(p)\,dp=\frac{1}{N\ell}R$. Then note that 
\begin{displaymath}
\begin{split}
N-N\ln N + N\left(\alpha+\beta\cdot \frac{1}{N\ell}R\right)=& N-N\ln N +N \int_{\mathbb{S}^2}\sigma(p)\left(\alpha +\beta \cdot p\right)\,dp\\
=&N-N\ln N +N\int_{\mathbb{S}^2}\sigma(p)\ln\left(\frac{\sigma(p)}{\rho^T(p)}\right)\,dp,
\end{split}
\end{displaymath}
so that the energy can be expressed in terms of the relative entroppy of $\sigma$ with respect to $\rho^T$. If we consider the problem of minimising 
$$\left(N-N\ln N +N\int_{\mathbb{S}^2}\tilde\sigma(p) \ln \frac{\tilde\sigma(p)}{\rho^T(p)}\,dp\right)$$
over all $\tilde\sigma \in\mathcal{P}(\mathbb{S}^2)$ such that
\begin{displaymath}
\begin{split}
\int_{\mathbb{S}^2}\tilde\sigma(p)\,dp=&1,\\
\int_{\mathbb{S}^2}p\tilde\sigma(p)\,dp=&\frac{1}{N\ell}R,
\end{split}
\end{displaymath}
we see that $\sigma(p)=\frac{1}{Z}\exp(\beta \cdot p)\rho^T(p)$ satisfies the Euler-Lagrange equation corresponding to the minimisation problem. This means that the derivation is equivalent to a maximum {\it relative} entropy assumption on the chain, whereas the classical Kuhn and Gr\"un model can be viewed as a maximum classical entropy assumption. It is this representation of the chain energy that will be most useful for the analysis.

\subsection*{Basic Properties}

By fixing $\rho$, and using the substitution $\rho^T\sigma'=\tilde\sigma$, the minimisation problem is equivalent to 
$$\min\limits_{\sigma' \in\mathcal{B}(R,\rho^T)} k_BT\left(N\ln N +N\int_{\mathbb{S}^2}\sigma'(p) \ln \sigma'(p)\rho^T(p)\,dp\right),$$
where $\mathcal{B}(R,\rho^T)$ is the set of all $\sigma':\mathbb{S}^2\to [0,+\infty)$ which are measurable with respect to $\rho^T(p)\,dp$, satisfying the linear constraints
\begin{displaymath}
\begin{split}
\int_{\mathbb{S}^2}\sigma'(p)\rho^T(p)\,dp=&1,\\
\int_{\mathbb{S}^2}p\sigma'(p)\rho^T(p)\,dp=&\frac{1}{N\ell}R.
\end{split}
\end{displaymath}
This is advantageous since this can be viewed as a classical entropy maximisation, but with the usual surface area measure replaced with $\rho^T(p)$. In particular, methods from \cite{borwein1991duality,taylor2015maximum} can be applied to immediately provide some relevant results. Recalling from Equation \eqref{eqRhoT} that $\rho^T=\frac{1-\gamma}{4\pi}+\gamma\rho$, where $\gamma$ is taken to be fixed, we note that the energy can be taken as dependent on $\rho$, the nematic orientation distribution function, rather than $\rho^T$. Define 
$$W_c(R,\rho)=\min\limits_{\sigma' \in\mathcal{B}(R,\rho^T)} k_BT\left(N\ln N +N\int_{\mathbb{S}^2}\sigma'(p) \ln \sigma'(p)\rho^T(p)\,dp\right).$$ 
Some key properties of $W_c$, which can be taken from results in the given reference, are that 
\begin{enumerate}
\item $W_c(R,\rho)$ is a smooth convex function of $R$ for $|R|<N\ell$.
\item For fixed $\rho$, $\beta$ is a smooth bijection between all $R \in \mathbb{R}^3$ with $|R|<N\ell$ and $\mathbb{R}^3$. 
\item As $|R|\to N\ell$, $W_c(R,N\ell)\to+\infty$, so the energy blows up in the taut chain limit. 
\item The value of $W_c$ can be found numerically by the dual problem
$$W_c(R,\rho)=k_BTN\ln N + k_BTN\max\limits_{(\alpha,\lambda) \in \mathbb{R}^{1+3}} \alpha+\frac{1}{N\ell}R\cdot \lambda - \int_{\mathbb{S}^2}\exp(\alpha-1+\lambda\cdot p)\rho^T(p)\,dp.$$
Furthermore the maximising pair $(\alpha,\lambda)$ has that $\lambda=\beta$.
\end{enumerate}

Two important properties, relating to invariance, are as follows.
\begin{enumerate}
\setcounter{enumi}{4}
\item $W_c$ is frame indifferent, so that if $S\rho$ is defined as $(S\rho)(p)=\rho(Sp)$ for a rotation $S$, then $W_c(SR,S\rho)=W_c(R,\rho)$. Similarly, $S\beta(SR,S\rho)=\beta(R,\rho)$. 
\item If $W_c^{N,\ell}$ explicitly denotes $W_c$ with given number of monomers $N$ of length $\ell$, 
$$\frac{1}{k_BTN}\bigg(W_c^{N,\ell}(R,\rho)-k_BTN\ln N\bigg)=W^{1,1}_c\left(\frac{1}{N\ell}R,\rho\right).$$
That is to say, up to an additive and multiplicative constant, $W_c$ depends only on $\frac{1}{N\ell}R$ and $\rho$. 
\end{enumerate}
These two invariance properties are readily seen by noting that the Euler-Lagrange equation for the minimisation problem is invariant under these transformations also. 

Property 3 demonstrates the finite extensibility of the chain. The blow up occurs because if $|R|\approx N\ell$, there is very little configuration space available. The more precise statement and argument can be found in \cite[Propositions 2.5 and 2.6]{taylor2015maximum}, and loosely states that if $|R|\approx N\ell$ then any probability distribution $\sigma\in\mathcal{P}(\mathbb{S}^2)$ that has $\frac{1}{N\ell}R$ as its first moment must be concentrated on some set with small size. This can be used to show that the energy, defined in terms of the relative entropy, must be large, with the precise argument found in \cite[Corollary 3.1]{taylor2015maximum}.

\section{Networks and the Quartic Approximation}

Take the head-to-tail symmetry assumption on $\rho$, so that $\rho(p)=\rho(-p)$ for all $p\in\mathbb{S}^2$. Using the formula from \cite[Appendix A]{taylor2015maximum}, the fifth order Taylor approximation of $W_c(\cdot,\rho)$ is given, up to an additive constant, as 
$$\frac{1}{Nk_BT}W_c^5(R,\rho)=\frac{1}{2(N\ell)^2}|V^\frac{1}{2}R|^2+\frac{1}{24(N\ell)^4}\left(3 L\otimes L-M\right):(VR)^{\otimes 4},$$
where $u^{\otimes i}$ denotes the tensor product of $u$ with itself $i$ times, and $A:B$ denotes $\sum\limits_{ijkl}A_{ijkl}B_{ijkl}$ for fourth order tensors, $L$ and $M$ are the second and fourth moment of $\rho^T$ respectively and $V=L^{-1}$. The form simplifies significantly from that given in \cite{taylor2015maximum} since all odd ordered moments of $\rho$ vanish due to the head-to-tail symmetry condition. 

Consider a polymer network, with cross-linking points $(y_i)_{i \in I}$. Take an affine displacement assumption, so that if the continuum deforms by an affine map $x\mapsto Fx+x_0$ for some matrix $F$ and vector $x_0$, then the cross-linking points also deform as $y_i \mapsto Fy_i+x_0$. This means that the end to end vectors $(R_{i,j})_{(i,j) \in I'}$, where $I'$ is a subset of $I\times I$, can be written as $R_{ij}=y_i-y_j$ therefore deform as $R_{i,j}\mapsto FR_{i,j}$. The affine-displacement assumption is common in models of isotropic elasticity (see e.g. \cite{boyce2000constitutive} or \cite{treloar1975physics}). This assumption in particular treats the chains as phantom, neglecting the effects of entanglements and volumetric effects. For this reason only incompressible systems will be considered. Furthermore we assume that the energy of deformation is simply the sum of of the energies of the particular chains. If the distribution of end-to-end vectors in the reference configuration is taken to be some probability measure $\nu$, and $n>0$ is the number of chains per unit volume, then the energy per unit volume is therefore given as 
$$W_D(F,\rho)=n\int_{\mathbb{R}^3}W_c(FR,\rho)\nu(R)\,dR.$$
At this point one could couple the energy of $\rho$ in one of two ways. Firstly by adding an Onsager-type contribution to the energy, so that for example
$$W_{\text{total}}(F,\rho)=W_D(F,\rho)+c_1\int_{\mathbb{S}^2} \rho(p)\ln\rho(p)-\rho(p) \int_{\mathbb{S}^2}K(p\cdot q)\rho(q)\,dq \, dp$$
for some appropriate constant $c_1$ and interaction kernel $K$. Alternatively, one could constrain $\rho$ to be in the minimising set for an Onsager-type energy, which follows the argument that the nematic contribution to the energy is greater than the elastic energy, and therefore can be minimised independently. The latter approach is many respects simpler since the minimising set is often simply generated by rotations acting on a particular solution, giving finitely many degrees of freedom in one-to-one correspondence with a subset of $\text{SO}(3)$. For the sake of this work though the contribution to the energy from the ``free" nematic is generally unimportant as we will only be considering $W_D$.

Substituting the Taylor expansion of the chain energy into the deformation energy thus gives, up to an additive constant,
\begin{displaymath}
\begin{split}
\frac{1}{k_BTNn}W_D^5(F,\rho)=\int_{\mathbb{R}^3}\left(\frac{1}{2(N\ell)^2}|V^\frac{1}{2}FR|^2+\frac{1}{24(N\ell)^4}\left(3 L\otimes L-M\right):(VFR)^{\otimes 4}\right)\,d\nu(R).
\end{split}
\end{displaymath}
Defining $L_0$ and $M_0$ to be the second and fourth order moments of $\nu$ respectively, and letting $\tilde{F}=VF$ for brevity, this integral can then be expanded as 
\begin{displaymath}
\begin{split}
&\frac{1}{2(N\ell)^2}|L^{\frac{1}{2}}\tilde{F}(L_0)^{\frac{1}{2}}|^2\\
&+\frac{1}{24(N\ell)^4}\left(3L_{i_1i_2}L_{i_3i_4}-M_{i_1i_2i_3i_4}\right)\tilde{F}_{i_1j_1}\tilde{F}_{i_2j_2}\tilde{F}_{i_3j_3}\tilde{F}_{i_4j_4} (M_0)_{j_1j_2j_3j_4},
\end{split}
\end{displaymath}
where the summation notation is used in the second line. This provides a quartic approximation, depending on the second and fourth moments of $\rho^T$ and $\nu$. 

This is similar to the expression given by Mao \cite{mao1998finite}, although in the work of Mao only a priori distributions $\rho^T(p)=\frac{!}{Z}\exp\left((\tilde{h}(p\cdot v)^2\right)$ for $n \in \mathbb{S}^2$ and $\tilde{h}\in\mathbb{R}$ are considered, which puts constraints on the relationships between $L$ and $M$ (the details of which are included in Appendix A of their work). The reader should take note of the difference in notation with the tensor $L_0$ in this work being proportional to $N\ell_0$ in the notation of Mao, so that the scaling with $N$ is consistent between the two works.

Furthermore, by truncating this to only second order the approximation can be written as 
$$\frac{1}{2(N\ell)^2}|L^{-\frac{1}{2}}FL_0^\frac{1}{2}|^2.$$ 
We see that up to a multiplicative constant, this is equivalent to the Neo-Hookean energy for a freely jointed main chain nematic elastomer (see \cite[Exercise 3.1]{warner2003liquid}), with the step tensor $L$ determined as the second moment of $\rho^T$. In particular this allows us to view the Neo-Hookean energy as a small-strain approximation to the energy presented in this work.

\subsubsection*{Network theories and symmetry}
If the chain energy is not replaced with its Taylor approximation then the behaviour will be more dependent on the precise form of $\nu$, rather than just some of its moments. In the examples of isotropic elasticity there are a large number of competing theories (see e.g. \cite{boyce2000constitutive} for a broad review) that correspond to using the KG energy with an affine displacement assumption, only varying the corresponding probability measure $\nu$. The requirement that the energy respects the material symmetry prevents immediate generalisations of many of these models, particularly the simpler models that take $\nu$ to have discrete support. 

To illustrate this issue, consider a nematic elastomer sample with uniaxial order in the reference configuration, about some axis $u$. We will take assumptions common to many models of isotropic elasticity. We assume that $\nu$ can be written as a sum of $k$ Dirac masses at $R_i=R_i(F,\rho)$, with $R_i$ respecting the material symmetry. Furthermore $|R_i(F,\rho)|=\ell\sqrt{N}$ for all $i=1,...,k$, $F\in\mathbb{M}^3$,$\rho\in\mathcal{P}(\mathbb{S}^2)$. The latter assumption corresponds to a root-mean-square assumption on the chains in the reference configuration. Then the energy can be written as
$$W_D(F,\rho;u)=n\sum\limits_{i=1}^k W_c\left(\frac{1}{N}FR_i(F,\rho),\rho\right).$$
If $F$ is a uniaxial extension/compression about $u$ then for all rotations $S$ with $Su=\pm u$, $SR_i(F,\rho)=R_i(FS^T,\rho)=R_i(F,\rho)$. There are however uncountably many of such rotations, so unless $R_i(F,\rho)= \ell\sqrt{N}u$ for all $i$, this cannot happen. In this case, deformations of the form $F=\frac{1}{\lambda^2}u\otimes u +\lambda\left(I-u\otimes u\right)$ for any $\lambda >0$ all have the same energy, removing any growth condition in the energy. This demonstrates that a discrete probability measure $\nu$ generally can't provide intuitive properties of the material.

Because of this issue relating to symmetry it may be the case that only the full-network models such as the Wu and Van der Geissen \cite{wu1993improved} model are the only type that admit a symmetry respecting generalisations. Their model considers $\nu$ to be supported on the sphere of radius $\ell \sqrt{N}$ isotropically, so that the energy is given by 
$$\int_{\ell\sqrt{N}\mathbb{S}^2} w_c\left(\frac{1}{N\ell}FR\right)\,dR,$$
and generalisations such as 
\begin{equation}\label{eqNetwork}
\int_{\ell\sqrt{N}\mathbb{S}^2}W_c\left(\frac{1}{N\ell}FR,\rho)\right)\,d\nu(R)
\end{equation}
may make appropriate candidates. 

We can illustrate the finitely extensible nature of the full network by considering a system isotropic at crosslinking, so that $d\nu(R)=\frac{1}{4\pi\ell^2N}dR$. In this case, if the largest singular value of the deformation gradient $F$ is larger than $\sqrt{N}$, then we have a set $A \subset \ell\sqrt{N}\mathbb{S}^2$ with $\nu(A)>0$ and $\left|\frac{1}{N\ell}FR\right|>N\ell$ for all $R \in A$. This implies that $\int_{\ell\sqrt{N}\mathbb{S}^2}W_c\left(\frac{1}{N\ell}FR,\rho\right)\,d\nu(R)=+\infty$, so that the arbitrary deformations are not permitted. Conversely, if the largest singular value of $F$ is strictly less than $\sqrt{N}$, then the energy is finite, so $\sqrt{N}$ as the extensibility limit just as in the Wu and Van der Geissen model from isotropic elasticity. 

The full-network model in isotropic elasticity is known for being both numerically and analytically intensive to work with, so we leave this open for future work. 

\section{Illustrative Figures}

For simplicity, only probability distributions of the form $\rho(p)=\frac{1}{z}\exp\left(a(p\cdot u)^2\right)$, where $a \in \mathbb{R}$ and $u \in \mathbb{S}^2$ will be considered for the nematic orientation distribution function. Probability distributions of this form are taken since they correspond to equilibria for the Maier-Saupe free energy \cite{fatkullin2005critical}, with the constant $a$ depending on the temperature. In each case we take $N\ell=1$ to provide a dimensionless version of the chain energy, and arbitrary units are taken. The relatively large number fraction $\gamma=0.3$ will be taken so that the behaviour is clearly visible. 

\begin{figure}[!htb]
\begin{subfigure}[t]{0.25\textwidth}
\includegraphics[width=0.95\textwidth]{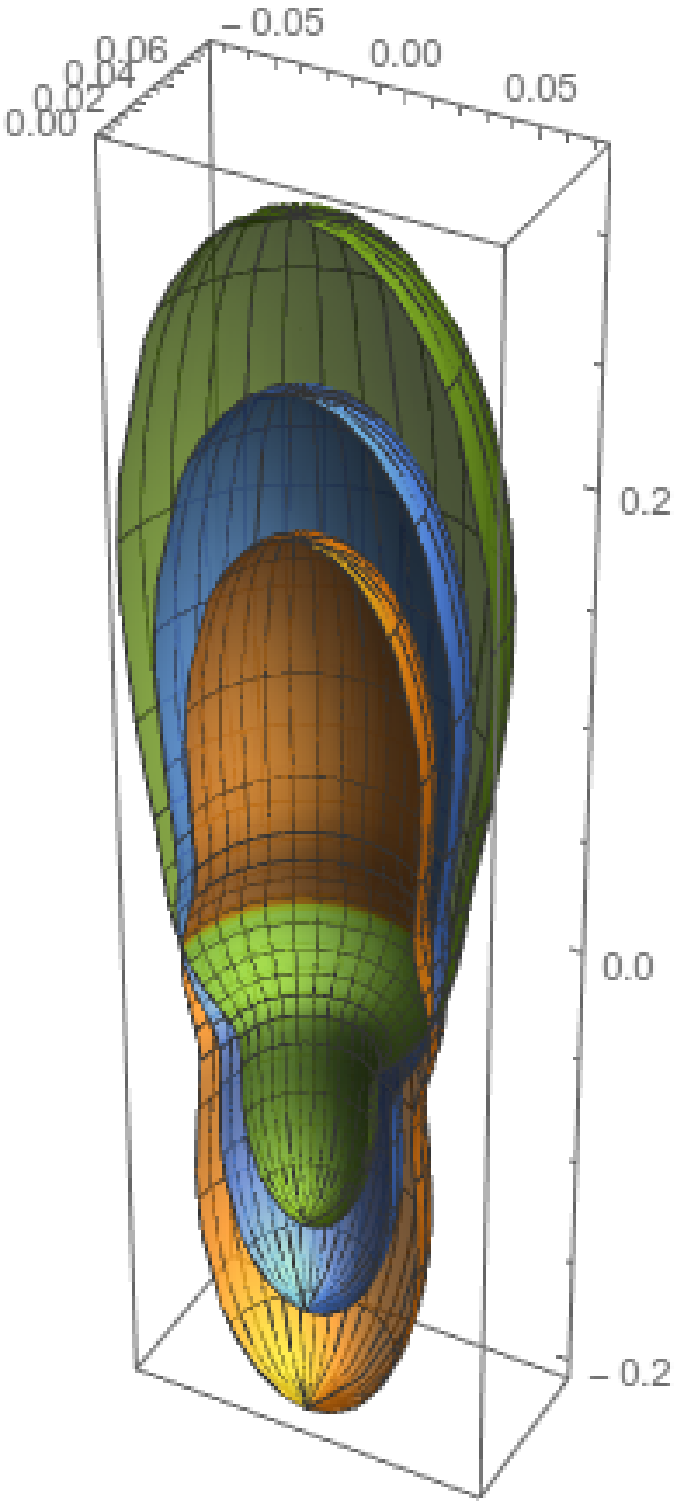} 
\caption{$\theta=0$}
\end{subfigure}
\begin{subfigure}[t]{0.25\textwidth}
\includegraphics[width=\textwidth]{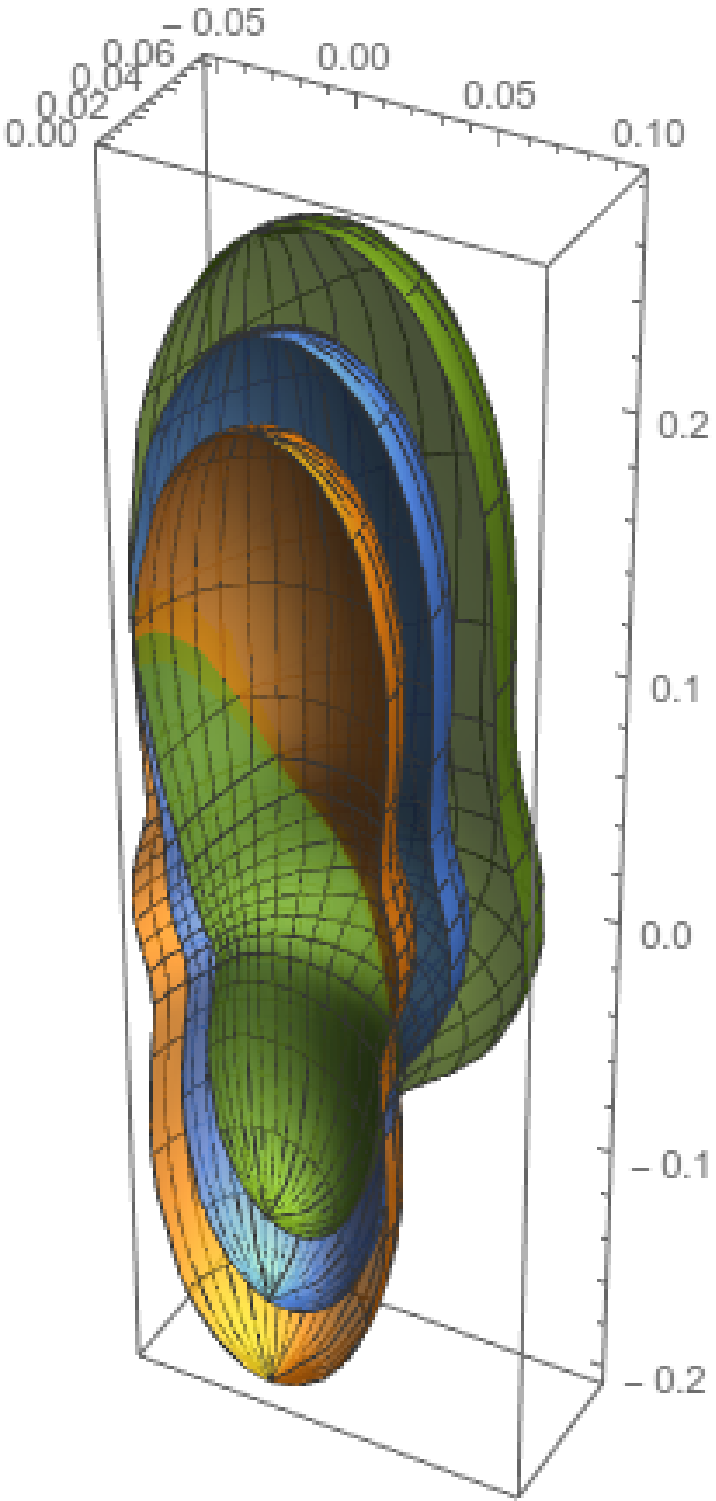} 
\caption{$\theta=\frac{\pi}{4}$}
\end{subfigure}
\begin{subfigure}[t]{0.4\textwidth}
\includegraphics[width=\textwidth]{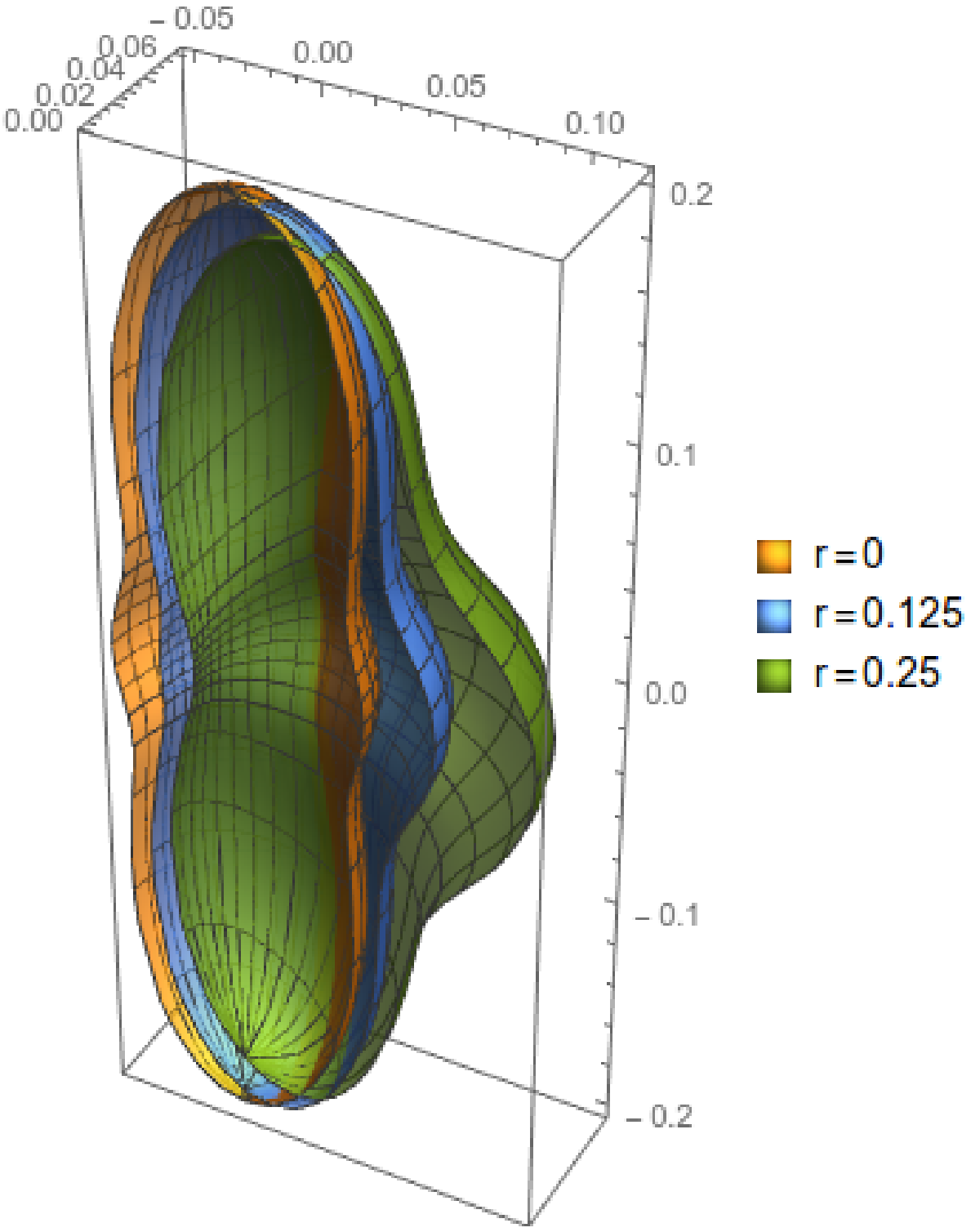} 
\caption{$\theta=\frac{\pi}{2}$}
\end{subfigure}
\caption{The optimal chain distribution $\sigma$ for $\frac{1}{N\ell}R=r\left(\cos(\theta)u+\sin(\theta)u^\perp\right)$.}
\label{figChainDistributions}
\end{figure}

In Figure \ref{figChainDistributions}, we take $\rho(p)=\frac{1}{Z}\exp(5(p\cdot u)^2)$. The corresponding Q-tensor for $\rho$ has distinguished eigenvalue of $\lambda \approx 0.431$. The unit vector $u^\perp$ is taken so that $u\cdot u^\perp=0$. The figure demonstrates the optimal orientation distribution $\sigma$ for various values of the nematic director and chain length. The three figures all show the expected symmetry, as well as elongation in the direction of $R$ as $\frac{1}{N\ell}|R|=r$ increases.

\begin{figure}[!htb]
\begin{center}
\includegraphics[width=0.5\textwidth]{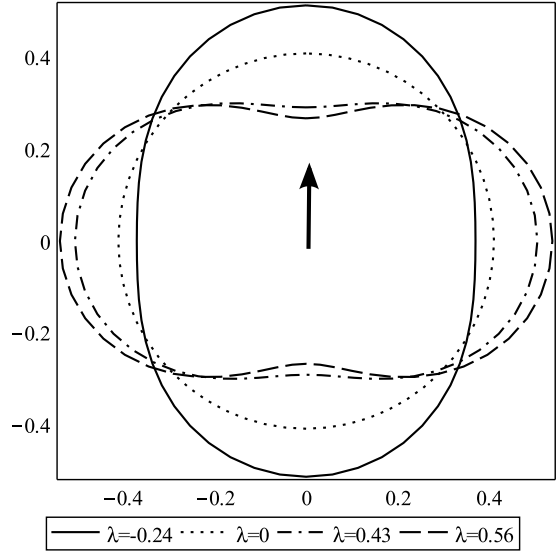}
\caption{$W_c\left(\frac{1}{2}e_\theta,\rho_\lambda\right)$, for various values of $\lambda$.}
\label{figWcLambdas}
\end{center}
\end{figure}

In Figure \ref{figWcLambdas} a plot in polar coordinates of $W_c\left(\frac{1}{2}e_\theta,\rho_\lambda\right)$ is shown, where $e_\theta=(\cos(\theta),\sin(\theta),0)$, and $\rho_\lambda$ is of the form given previously with its corresponding Q-tensor having distinguished eigenvalue $\lambda$, and corresponding eigenvector $n=e_0$. The coordinate direction $e_0$ is shown by the arrow on the figure. The figure illustrates that in a prolate nematic phase, stronger alignment of the nematic gives stronger preference to the chain aligning parallel to the director, and with oblate nematics there is an energetic preference for the chain to be perpendicular to the director. The heuristics to understand why this happens is that, particularly for larger $|R|$, any $\sigma$ such that $\int_{\mathbb{S}^2}p\sigma(p) \,dp = \frac{1}{N\ell}R$ must be large in the region where $p\approx \frac{R}{|R|}$ (see \cite[Proposition 2.5]{taylor2015maximum}). In the KG-model this would give a high energy, however in the anisotropic model the energy can be lowered if $\rho^T(p)$ is also large in the same region. This suggests for $R$ with fixed length but variable orientation, $W_c(R,\rho)$ will be lower in regions where $\rho^T\left(\frac{R}{|R|}\right)$ is larger. 

\begin{figure}[!htb]
\begin{center}
\begin{subfigure}[t]{0.45\textwidth}
\includegraphics[width=\textwidth]{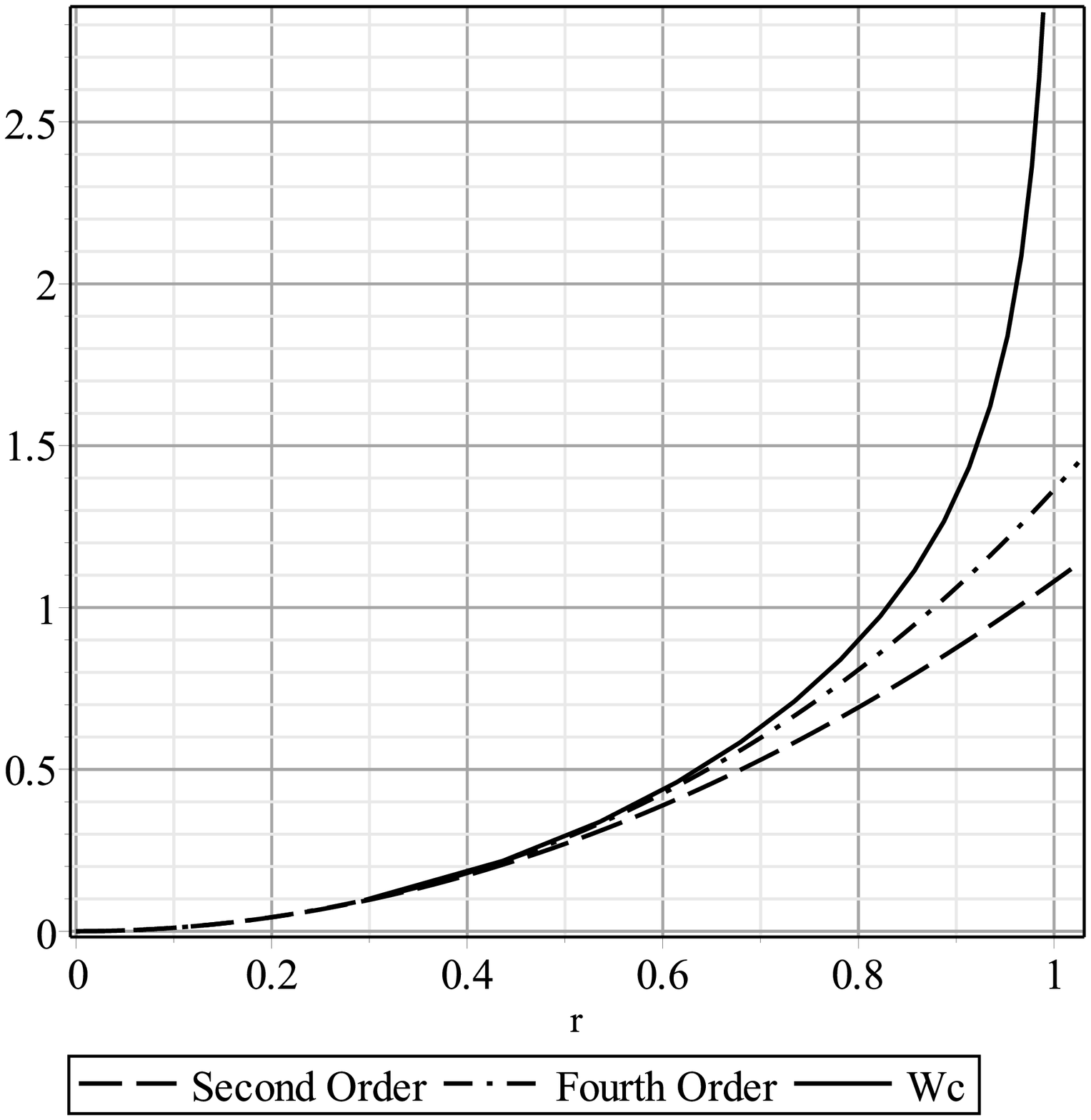} 
\caption{Parallel alignment}
\end{subfigure}
\begin{subfigure}[t]{0.45\textwidth}
\includegraphics[width=\textwidth]{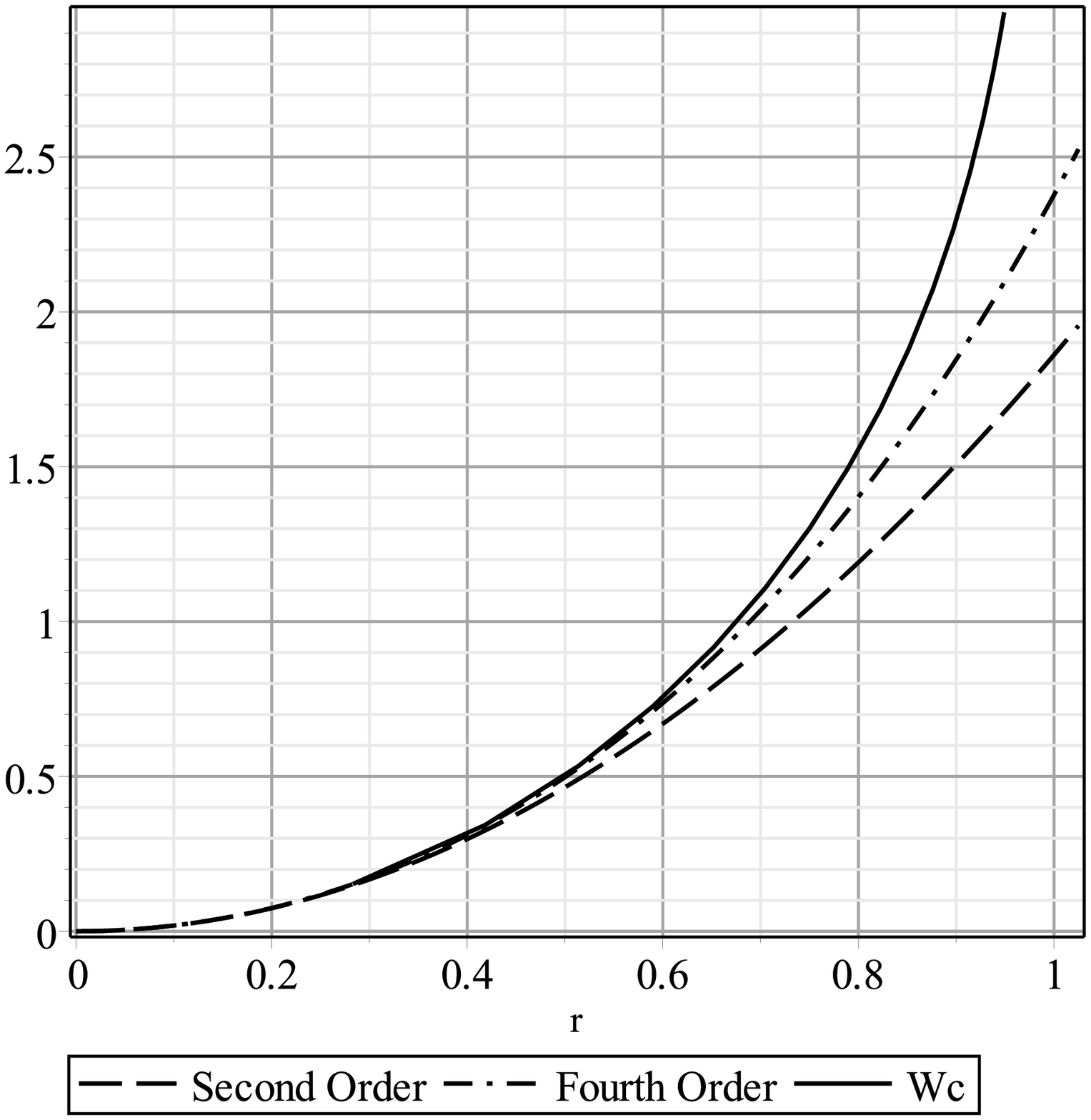} 
\caption{Perpendicular alignment}
\end{subfigure}
\caption{The ``full" finitely extensible chain energy and its approximations with parallel and perpendicular alignment.}
\label{figWcAndApprox}
\end{center}
\end{figure}

In Figure \ref{figWcAndApprox} again $\rho(p)=\frac{1}{Z}\exp(5(p\cdot u)^2)$ is taken. The figure shows $r=|R|$ against the energy, with $\hat{R}$ either parallel or perpendicular to the nematic alignment. The asymptote at $|R|=1$ is shown in dashed grey, and the second and fourth Taylor approximations are also included for comparison. Just as in the Neo-Hookean model, the true chain energy and its approximations show that the energetic preference is for the chain to be aligned with the nematic. Furthermore, similarly to the case in isotropic elasticity (see \cite{treloar1975physics}) it can be seen that the second order approximation is accurate up to around a third of the maximum extension.

\section*{Acknowledgements}
The author would like to thank John Ball for ongoing discussion that lead to the results in this work, as well as Maria-Carme Calderer and Peter Palffy-Muhoray for providing insights in our discussions, and to the organisers of the MLC program at the Isaac Newton institute. The research leading to these results has received funding from the European Research Council under the European Union's Seventh Framework Programme (FP7/2007-2013) / ERC grant agreement n$^{\circ}$ 291053.

\bibliographystyle{ieeetr}
\bibliography{bibligraphy}
\end{document}